\documentclass[prb,aps,twocolumn,amsmath,amssymb,floatfix,superscriptaddress]{revtex4}

\usepackage{color}
\usepackage{soul}
\usepackage[colorlinks=true, citecolor=blue, urlcolor=blue]{hyperref}
\usepackage{graphics,epstopdf}
\usepackage{epsf,graphics,graphicx}
\usepackage{float}
\begin{document}

\title{Quantum charge pumping in helical systems: A comparative study of short- and long-range hopping}

\author{Leila Eslami}
\email{eslami.leila1983@gmail.com; l.eslami@srbiau.ac.ir}
\affiliation{Department of Physics, SR. C., Islamic Azad University, Tehran, Iran}

\author{Santanu K. Maiti}
\email{santanu.maiti@isical.ac.in}
\affiliation{Physics and Applied Mathematics Unit, Indian Statistical Institute, 203 Barrackpore Trunk Road, Kolkata-700 108, India}

\author{Fatemeh Bourbour}
\affiliation{Department of Physics Education, Farhangian University, P.O. Box 889-14665 Tehran, Iran}

\date{\today}

\begin{abstract}
	
Using the Keldysh non-equilibrium Green's function approach, we investigate charge pumping through a single-stranded helical structure 
described by a tight-binding model that includes either short-range hopping (SRH) or long-range hopping (LRH). While quantum pumping has 
been studied in various low-dimensional systems, the detailed behavior of the spectral current and the pumped dc current in helical 
geometries in the presence of higher-order electron hopping (beyond nearest neighbors) has not yet been systematically explored. Here, 
we focus on the interplay between helicity and extended hopping ranges, analyzing how they jointly control the energy-resolved and dc 
pumped currents under time-periodic end potentials. For LRH, the pumped dc current exhibits pronounced plateau-like regions as a function 
of chemical potential when energy levels are sparsely spaced--consistent with adiabatic transport--whereas SRH yields more parameter-sensitive
currents without clear plateaus. The plateau stability is controlled by the drive frequency: at higher frequencies, Floquet side-band mixing
destroys the plateaus, leading to oscillatory currents. The phase dependence remains nearly sinusoidal, and the current vanishes at zero 
phase lag, confirming the necessity of out-of-phase potentials. Crucially, in helical systems, the decay exponent $(\ell_c)$ acts as an effective structural parameter that can tune both the magnitude and sign of the pumped current, offering a geometric knob for controlling quantum pumping. Our findings not only fill a gap in the understanding of spectral and pumped currents in helical systems with extended 
hopping but also provide tools that can be applied to analyze similar phenomena in other chiral or quasi-one-dimensional systems.

\end{abstract}

\maketitle

\section{Introduction}
\label{sec:introduction}

Coherent control of charge transfer at the nanoscale plays a key role in molecular electronics and quantum enabled technologies~\cite{nitzan2003electron, Datta1995,Awschalom2018,ESLAMI2016,Farshchi2020,Torres2022,Torres2024}. Among the diverse platforms explored, molecular structures with helical geometries--such as DNA and $\alpha$-helical proteins--have emerged as promising candidates 
due to their intrinsic chirality~\cite{Sarkar2019,Constantinos2023,Constantinos2024,Aminiranjabar2025,Constantinos2025}, tunable electronic
properties~\cite{Aminiranjabar2025,Chen2023}, and self-assembly capabilities~\cite{Aminiranjabar2025}.

Quantum pumping~\cite{Switkes1999}--the generation of a direct current, in the absence of an explicit bias voltage, through the cyclic modulation of system parameters--represents a fundamental concept of time-dependent quantum transport. Since the pioneering work of Thouless~\cite{Thouless1983}, charge pumping has been extensively studied in mesoscopic systems such as quantum dots~\cite{Quantumdotpump,Haughian2017,Monsel2022}, nanowires~\cite{Faizabadi_2004,Arrachea2005,DalLago2015}, molecular junctions~\cite{LRH2020,Zhou2023,TuovinenPavlyukh2024}, and two-dimensional materials~\cite{Satofumi2014,Ridley2017,Bourbour2020,Nikoofard2022}. The theoretical framework, often built on the scattering-matrix or non-equilibrium Green's function (NEGF) formalism, is well established~\cite{Arrachea2007,Imry1999}. Recent reviews provide a comprehensive overview of the underlying mechanisms, including 
geometric effects and non-adiabatic photon-assisted tunneling~\cite{Acciai2025}. Yet, these studies predominantly focus on 
low-dimensional models where electron hopping is typically restricted to nearest neighbors.

In a quantum pumping nanojunction, such as the system considered in this work, a DC current is generated solely by time-periodic potentials,
without an external bias voltage. Consequently, the transport mechanism fundamentally differs from that of conventional bias-driven nanojunctions: since no bias contacts are required, energy dissipation associated with contact resistance is intrinsically 
suppressed~\cite{Marchesoni2009,Lee2018}. This bias-free operation represents a key advantage of pumping-based devices, particularly for low-power and nanoscale applications. Beyond the absence of dissipation, pumped currents also differ qualitatively from conventional 
currents in terms of the electronic states contributing to transport. In standard bias-induced transport, only electrons within the bias 
window participate. In contrast, in pumping transport, depending on the driving frequency and amplitude, electrons over a broad range 
of energies can participate, reflecting the intrinsically non-equilibrium and time-dependent nature of quantum pumping.

Another critical difference is how the current direction is determined. In conventional transport, the current direction is uniquely 
determined by the sign of the applied bias voltage. In pumping transport, however, the direction of the DC current is governed by the 
imbalance between the probabilities for electron tunneling from left to right and from right to left~\cite{Arrachea2007}. Importantly, 
electrons at different energies may contribute currents with opposite directions, and the sign of the net pumped current emerges only 
after summing over all energy-resolved contributions. As a result, the magnitude and even the direction of the pumped current cannot, 
in general, be inferred from simple a priori considerations, underscoring the complex nature of electron dynamics in driven quantum 
systems. Concurrently, significant research has been devoted to understanding the electronic properties of helical molecules in the 
steady-state regime~\cite{Sarkar2019,Chen2023,Aminiranjabar2025}. For instance, in a single-stranded magnetic helix subjected to a 
perpendicular electric field, the interplay between chirality and helical geometry--controlled by parameters such as radius, pitch, 
and decay constant--strongly modulates spin-dependent transmission spectra~\cite{Sarkar2019}. 
\begin{figure}[ht]  
{\centering \resizebox*{6cm}{10cm}{\includegraphics{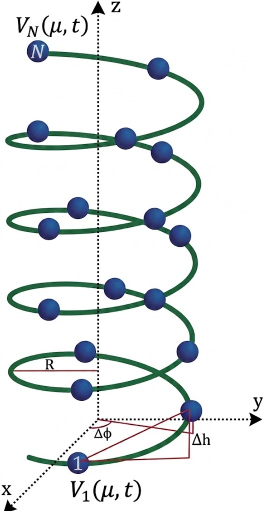}}\par}
\caption{(Color online). Schematic illustration of a right-handed single-stranded helical molecule sandwiched between two reservoirs kept 
at the same chemical potential, as a pumping device. $R$ represents the radius, $\Delta \phi$ is the twisting angle, and $\Delta h$ denotes 
the stacking distance between the neighboring lattice sites.}
\label{fig:1}  
\end{figure}
Moreover, this work demonstrates that the transition from short-range hopping to long-range hopping regimes dramatically alters the 
electronic spectrum, leading to level clustering and enhanced delocalization. Another recent complementary experimental-theoretical study 
on DNA duplexes has shown that nearest-neighbor base-pair interactions play a crucial role in charge delocalization and coherence length~\cite{Aminiranjabar2025}. By designing DNA sequences, the electronic density of states can be engineered to maintain high 
conductance over lengths of up to $20$ base pairs ($\approx 7\,$nm). This work provides explicit design guidelines for controlling 
charge transport in helical bio-molecules, highlighting how sequence-specific coupling can be used to optimize coherent transport.

Despite these advances in understanding the static electronic structure of helices, the exploration of time-dependent quantum phenomena,
particularly quantum charge pumping, in helical architectures has rarely been studied~\cite{LRH2020,Zhou2023}. An important development 
was made by Guo et al.~\cite{LRH2020}, who demonstrated that Thouless pumping in helical molecules is tunable via long-range hopping interactions, leading to a topological phase transition with reversible quantized current. Their findings highlight the potential of 
biological helices as candidate materials for designing molecular electronic devices. In another theoretical study published in 2023, 
Zhou and Kanai explored non-adiabatic Thouless pumping of electrons using the approach of topological Floquet engineering, with a 
particular focus on how chemical modifications can govern the emergence of the Floquet topological phase~\cite{Zhou2023}. By employing 
real-time time-dependent density functional theory (RT-TDDFT), the study examines the influence of molecular-level substitutions on the topological invariant to explain the interplay between chemical structure and quantum transport in driven systems. However, despite these advances, the detailed spectral current and pumped DC current in helical systems with higher-order electron hopping (beyond nearest neighbors) have not been systematically explored. In particular, the interplay between helicity and extended hopping ranges, and its effect on the 
current spectral function and DC pumped currents, remains an open question, which we address in this work.

We investigate the spectral currents and pumped DC currents in a single-stranded helical molecule, comparing short-range hopping and 
long-range hopping configurations using the Keldysh NEGF formalism. We focus on the interplay between helicity and higher-order hopping, 
an aspect not discussed in detail before. Our analysis covers both adiabatic and non-adiabatic regimes. We show that LRH exhibits 
plateau-like currents for sparse energy levels in low frequencies, while higher frequencies destroy these plateaus. Moreover, the decay 
exponent $\ell_c$ can tune both the magnitude and sign of the pumped current. Our results establish helical molecules with long-range 
hopping as promising platforms for efficient quantum charge pumps.

The paper is organized as follows. In Sec. II, we describe the tight-binding model for the helical system and detail the Keldysh NEGF
methodology for time-dependent transport. Section III presents our numerical results, comparing the spectral current, pumped dc current, 
and its dependence on chemical potential, phase lag, driving amplitude, frequency, and helical decay constant for both SRH and LRH regimes. 
Section IV includes a summary and an outlook.

\section{Quantum system and the Method}

\subsection{Tight-binding model for the helical structure}
 
In this study we have investigated electron pumped current properties in a DNA-like helical molecule. The system under consideration is 
a right-handed single-stranded helical molecule sandwiched between two reservoirs kept at the same chemical potential ($\mu$). To confine 
the system, potential barriers are introduced at both ends of the helix. By discretizing the energy bands, the barriers give rise to 
Fabry–Pérot oscillations resulting from the interference of different states within the region confined between the two static barriers~\cite{PhysRevB.77.085408, PhysRevB.77.153301}.
Additionally, two out of phase time-dependent periodic potentials are applied to the ends of the central region, as shown in 
Fig.~\ref{fig:1}. According to the figure, $\Delta \phi$, $\Delta h$ are the twisting angle and stacking distance between two consecutive
lattice sites, respectively.  
Depending on the value of $\Delta h$, two distinct configurations of helical molecules are considered: the long-range hopping (LRH) regime and the short-range hopping (SRH) regime. For small values of the stacking distance $\Delta h$, electron hopping occurs not only between nearest-neighbor sites but also among more distant sites along the helical molecule. This situation corresponds to the LRH configuration. In contrast, for large values of $\Delta h$, the overlap between distant sites becomes negligible, and hopping is predominantly restricted to nearest neighbors. The system is then said to be in the SRH regime. To describe the time-dependent helical molecule, we employ a tight-binding (TB) framework. The system is described by the Hamiltonian
\begin{equation}
H =H_{helix}+H_{res}+H_{cpl}+\sum_{i=1, N} V_p \cos (\Omega t+\delta_{i}),
\label{eq:totalH}
\end{equation}
where $H_{helix}$, $H_{res}$, and $H_{cpl}$ are the Hamiltonian of the helix, the reservoirs, and the coupling between the reservoirs and the central system, respectively. The last term in Eq.~\ref{eq:totalH} represents the time-dependent potentials applied on the two ends of the helix. Here, $V_p$ and $\Omega$ show the amplitude and frequency of the driving potential, respectively. Moreover, $\delta_{i}$ is the 
phase of the time-dependent potentials with $i=1, N$ representing the bottom and top potentials, respectively and the phase lag is defined 
as $\delta=\delta_{N}- \delta_1$. The TB Hamiltonian for an $N$-site helix, in both SRH and LRH regimes, is expressed as~\cite{Guo2014}
\begin{equation}
H_{helix} =\sum_{i=1}^{N}\epsilon_i c_i^\dagger c_i + \sum_{i=1}^{N-1} \sum_{j=1}^{N - i} 
\left( t_j\, c_i^\dagger c_{i+j} + h.c.\right),
\end{equation}
where $\epsilon_i$ denotes the on-site energy at site $i$, and $c_i^\dagger$ ($c_i$) is the fermionic creation (annihilation) operator at 
that site. In this work, the on-site energies are set to zero for all sites except at the two ends of the helix ($i = 1, N$), where 
potential barriers of height $E_b$ are introduced. Accordingly, we take $\epsilon_i = 0$ for $i = 2, \dots, N-1$, and 
$\epsilon_1 = \epsilon_N = E_b$. The parameter $t_j$ represents the hopping amplitude between sites $i$ and $i+j$, and its form can 
be expressed as~\cite{Sarkar_2020}
\begin{equation}
t_j=t_0 \exp(-(l_j-l_1)/l_c),
\end{equation}
where $t_0$ is the nearest-neighbor hopping (NNH) strength, $l_j$ is the separation distance between the sites $i$ and $i+j$, and $l_c$ 
is the decay constant. The separation distance $l_j$ can be expressed in terms of the helix structural parameters as~\cite{Sarkar_2020}
\begin{equation}
l_j = \sqrt{ \left[ 2R \sin\left( \frac{j \Delta \phi}{2} \right) \right]^2 + (j \Delta h)^2 },
\end{equation}
where $R$ is the radius of the helix. The Hamiltonian $H_{res}$ representing external electrodes connected to the Helix can be written as
\begin{equation}
H_{res} = \sum_{i,\alpha=t, b}  \epsilon_0 a_{i,\alpha}^\dagger a_{i,\alpha} + \sum_{\langle i,j \rangle, \alpha=t, b} (t_r a_{i,\alpha}^\dagger  a_{j,\alpha} + \text{H.c.}),
\end{equation}
where $a_{i,\alpha}^\dagger$ ($a_{i,\alpha}$) is the fermionic creation (annihilation) operator at site $i$ at the reservoir 
$\alpha=t, b$ with $t$ and $b$ standing for top and bottom leads, respectively. The parameter $t_r$ represents the hopping amplitude 
inside the reservoirs.
Moreover, the reservoirs are coupled to the end points of the helical system. The coupling Hamiltonian ($H_{cpl}$) is given by
\begin{equation}
 H_{cpl} = -w_b(a_b^{\dagger}c_1+ h.c.)-w_t(a_t^{\dagger}c_N +  h.c.),
\end{equation}
where $w_b (w_t)$ is the coupling constant between the bottom (top) reservoir and first $(Nth)$ site on the helix.   

The escape rate of electrons from the central region into leads $\alpha=t, b$ is given by the factor 
$\Gamma_{\alpha}(\omega)=i \bigl[ \Sigma_\alpha(\omega) - \Sigma_\alpha^\dagger(\omega) \bigr]$, which describes the coupling between the central region and lead $\alpha$. Here, $\Sigma_\alpha(\omega)$ is the self-energy arising from lead $\alpha$, 
which can be obtained using the recursive Green's function method~\cite{LopezSancho1985}.

\subsection{Keldysh non-equilibrium Green's function formalism}

In what follows, we use Keldysh non-equilibrium Green’s function formalism (NEGF) to study the time evolution of quantum states of 
the open quantum system and calculate the pumped charge current. According to the literature~\cite{Rammer1986}, the retarded Green's 
function between two arbitrary sites ($m$ and $n$) of the system is
\begin{equation}
\begin{aligned}
G^{R}_{m,n}(t,t') &= -i\Theta(t-t')\langle \{c_m(t),c_{n}^\dagger(t')\} \rangle.\\
\end{aligned}
\end{equation}
The integral representation of the Dyson equation for the Green's functions is~\cite{Arrachea2006}
\begin{widetext}
\begin{align}
G^{R}_{m,n}(t,t') &= G^{0}_{m,n}(t-t') + 
\sum_{j=1,N}\int dt_{1}\,
G^{R}_{m,j}(t,t_{1})V_{j}(t_{1})
G^{0}_{j,n}(t_{1},t'),
\end{align}
\end{widetext}
where $V_{j}(t)$ is the time periodic voltage applied at positions with co-ordinate $j=1,N$ which are connected to the reservoirs. 
Here, the unperturbed retarded Green's function $G^0_{m,n}(t-t')$ describes the central system in the absence of time-dependent 
potentials. Using the partial Fourier transform with respect to the difference time, $(t-t')$, the Green's function is expressed 
in the mixed time-frequency domain as
\begin{align}
G^{R}_{m,n}(t,\omega) &= \int_{-\infty}^{t} \! dt'\,
e^{i\omega(t-t')} G^{R}_{m,n}(t,t').
\end{align}
This quantity $G^{R}_{m,n}(t,\omega)$ is periodic in $t$ (with period $T_0=\frac{2\pi}{\Omega}$) and will be used in the following 
to derive the Dyson equation in the Floquet representation. For the retarded Green's function a set of Dyson equations can be 
written as~\cite{Arrachea2005}
\begin{align}
\hat{G}^{R}(t,\omega)
&= \hat{G}^{0}(\omega)  \nonumber\\[4pt]
&\quad+ \sum_{|n|\ge 1}
    e^{-in\Omega t}\,
    \hat{G}^{R}(t,\omega+n\Omega)
    \hat{V}(n\Omega)
    \hat{G}^{0}(\omega),
\label{eq:Dyson}
\end{align}
where $\hat{G}^{R}(t,\omega)$ and $\hat{G}^{0}(\omega)$ are $N \times N$ matrices with elements $G^{R}_{m,n}(t,\omega)$ and $G^{0}_{m,n}(\omega)$, respectively. $\hat{V}(n\Omega)$ is the perturbation matrix corresponding to the time periodic voltages. 
For the single-harmonic drive $[V(t)=V_p\cos(\Omega t+\delta)]$ the only non-zero components are $V(\pm \Omega)=(V_p/2)e^{\mp i\delta}$. 
By considering single harmonic potential case where $\hat {V}(n\Omega)=0$ for $\left| n \right| \ge 2$, Eq.~(\ref{eq:Dyson}) reduces to 
\begin{align}
\hat{G}^{R}(t,\omega)
&= \hat{G}^{0}(\omega)  \nonumber\\[4pt]
&\quad+
    e^{-i\Omega t}\,
    \hat{G}^{R}(t,\omega+\Omega)
    \hat{V}(\Omega)
    \hat{G}^{0}(\omega)\nonumber\\[4pt]
   &\quad+
       e^{i\Omega t}\,
    \hat{G}^{R}(t,\omega-\Omega)
    \hat{V}(-\Omega)
    \hat{G}^{0}(\omega).
\label{eq:Dyson_reduced}
\end{align}
Equation~(\ref{eq:Dyson_reduced}) shows the coupling of the electrons to the periodic external field. Here, the total Green’s function 
is obtained by considering the contributions from the unperturbed system as well as processes involving energy exchange of $\pm \Omega$ (photon-assisted absorption and emission) while the electron passes through the conductor. Using the Fourier series expansion of the 
solutions of Eq.~(\ref{eq:Dyson_reduced}) we get,     
\begin{equation}
\hat{G}^{R}(t, \omega) = \sum_{m = -K}^{K} \hat{\mathcal{G}}(m, \omega) e^{-i m \Omega t},
\label{eq:Fourier}
\end{equation}
which results in a change of representation from time to Floquet-frequency (harmonic) representation, where $\hat {\mathcal{G}}(m,\omega)$ 
is the Floquet component defined by $\hat{\mathcal{G}}(m,\omega) = \frac{1}{T_0} \int_0^{T_0} 
\hat {G}^R(t,\omega) e^{im\Omega t} dt$.

The final results for the Dyson's equation are~\cite{Arrachea2006}

\noindent\textbullet\ for $m=0$
\begin{equation}
\begin{aligned}
\hat{\mathcal{G}}(0, \omega)
= \Big[ & (\hat{G}^0(\omega))^{-1}
- \hat{V}(+\Omega)\hat{g}^{(+1)}(\omega+\Omega)\hat{V}(-\Omega) \\
& - \hat{V}(-\Omega)\hat{g}^{(-1)}(\omega-\Omega)\hat{V}(+\Omega)
\Big]^{-1},
\end{aligned}
\end{equation}
\noindent\textbullet\ for  $m> 0$
\begin{equation}
\hat{\mathcal{G}}(m,\omega)
=
\hat{g}^{(+m)}(\omega+m\Omega)\,
\hat{V}(-\Omega)\hat{\mathcal{G}}(m-1,\omega),
\end{equation}
\noindent\textbullet\ for $m < 0$
\begin{equation}
\hat{\mathcal{G}}(m,\omega)
=
\hat{g}^{(-|m|)}(\omega+m\Omega)\,
\hat{V}(+\Omega)\hat{\mathcal{G}}(m+1,\omega).
\end{equation}
Here, $\hat{g}^{\pm m}(\omega, \pm m\Omega)$ can be obtained by a recursive procedure: 
\begin{multline}
\big[\hat{g}^{(\pm m)}(\omega \pm m \Omega)\big]^{-1}
= \big[\hat{G}^0(\omega \pm m \Omega)\big]^{-1} \\
- \hat{V}(\pm \Omega)\,
  \hat{g}^{\pm m \pm 1}(\omega \pm (m+1)\Omega) \\
\times \hat{V}(\mp \Omega),
\end{multline}
considering that there is a finite cut-off integer $K$ such that $\hat{g}^{\pm K}(\omega \pm K\Omega)=\hat{G}^0(\omega \pm K\Omega)$ 
where $K$ is sufficiently large to ensure numerical convergence.

\subsection{Spectral current and pumped dc current}

The current spectral function (i.e., current per unit energy) is obtained, using the renormalization prescription of the previous 
section, in terms of the Floquet components of the retarded Green's function $(\hat{\mathcal{G}}(n,\omega))$~\cite{Arrachea2006}
\begin{multline}
\hat{j}_{\alpha}(\omega)
= 2\,\operatorname{Im}\!\left[\hat{\mathcal{G}}(0,\omega)\right]
  \hat{\Gamma}_{\alpha}(\omega) f_{\alpha}(\omega) \\
+ \sum_{\beta=t, b} \sum_{\kappa=-K}^{K}
\hat{\Gamma}_{\alpha}(\omega+\kappa \Omega)\,
f_{\beta}(\omega)\,
\hat{\mathcal{G}}(\kappa ,\omega)
\hat{\Gamma}_{\beta}(\omega)
\hat{\mathcal{G}}^{\dagger}(\kappa ,\omega).
\end{multline}
Here, $\hat{\Gamma}_{\alpha (\beta)}(\omega)$ denotes the hybridization matrix, and $f_{\alpha (\beta)}(\omega)$ is the Fermi distribution function of the contact $\alpha (\beta) = t , b$.

Finally, for a system driven by a time periodic potential with period $T_0$ the dc component of the current is defined as
\begin{equation}
J_\alpha^{dc}=\frac{1}{2\pi}{\int_{-\infty}^{\infty}d\omega j_\alpha (\omega)},
\end{equation}
where $j_\alpha (\omega)=\text{Tr}[\hat{j}_\alpha (\omega)]$.

\begin{table}[t]
\centering
\caption{Physical parameters of the right-handed LRH and SRH helices.}
\label{tab:helix_params}
\begin{ruledtabular}
\begin{tabular}{ccccc}
Helix & $R$ (\AA) & $\Delta h$ (\AA) & $\Delta \phi$ (rad) & $l_c$ (\AA) \\
\hline
SRH & 7   & 3.5 & $\pi/5$     & 0.9 \\
LRH & 2.5 & 1.5 & $5\pi/9$    & 0.9 \\
\end{tabular}
\end{ruledtabular}
\end{table}
 
\section{Numerical Results and Discussion}

This section presents the numerical results for the pumped charge current obtained using the Keldysh non-equilibrium Green's function 
formalism described in the previous section. Throughout this work, $t_0$ is taken as the unit of energy, and all other parameters are 
scaled accordingly. Upon restoring the units of $\hbar$, the resulting current is expressed in units of $e[E]/\hbar$, where $e$ denotes 
the electronic charge, $\hbar$ is the reduced Planck constant, and $[E]$ represents the unit of energy. For $[E]$ measured in $\text{eV}$, 
the current is found to be of the order of $10^{-4},\text{A}$. To improve the readability of the plots, the current has been multiplied 
by a factor of $100$, and hence the current axis is presented in units of $\mu\text{A}$. Unless otherwise stated, the structural parameters 
for both the long-range and short-range hopping models are adopted from the literature~\cite{PhysRevB.82.125125} and are summarized in Table I.

\begin{figure}[ht]
{\centering\resizebox*{8cm}{5cm}{\includegraphics{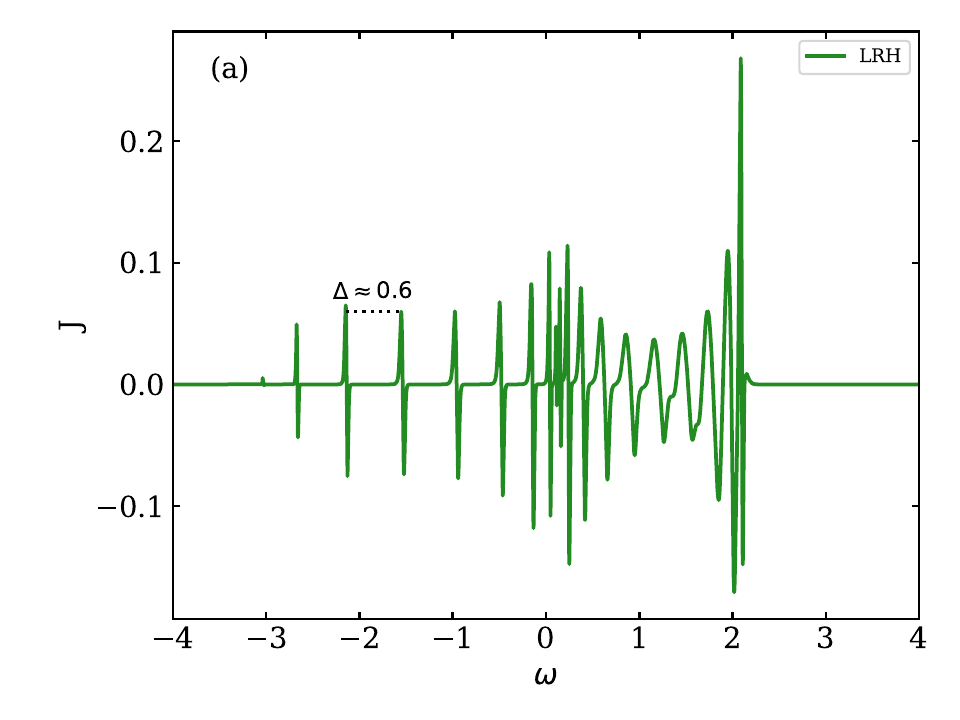}}
\resizebox*{8cm}{5cm}{\includegraphics{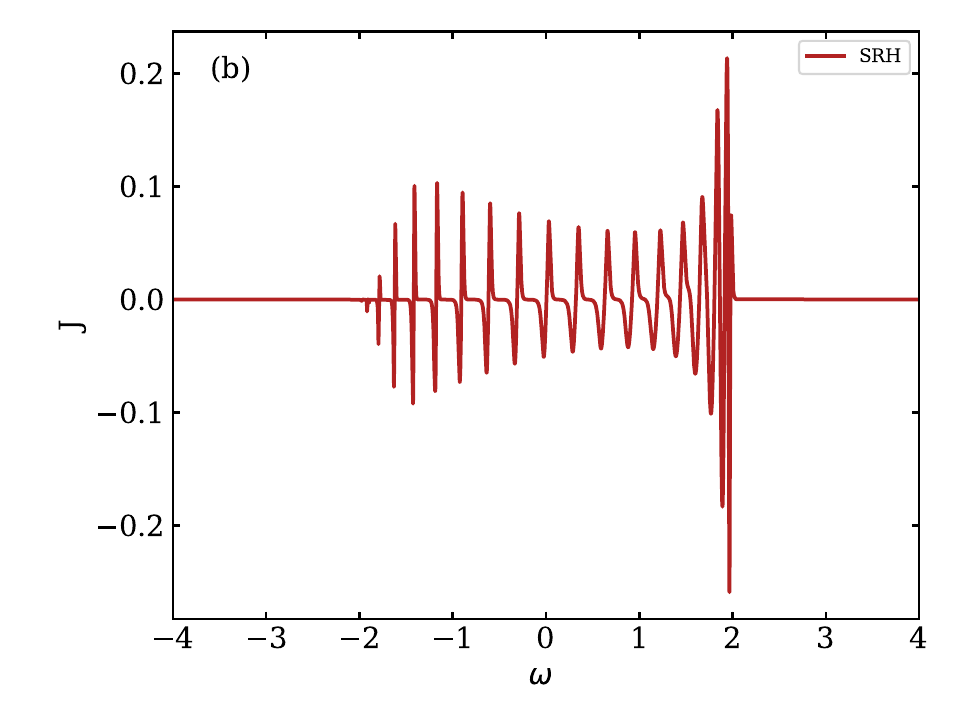}}\par}
\caption{(Color online). The current spectral function versus the electron energy for (a) LRH and (b) SRH configurations with $V_p= 0.5$ 
and $\Omega = 0.01$.}
\label{fig:2}
\end{figure}
The number of sites on the helix is $N=20$ for the LRH and SRH cases. Here, the selected parameters correspond to those of DNA and $\alpha$-helical protein molecules, which are well-known examples of SRH and LRH models in the literature~\cite{Endres2004}.

Figures~\ref{fig:2}(a) and~\ref{fig:2}(b) illustrate the current spectral function $(J)$ versus electron energy ($\omega$) for the 
LRH and SRH configurations, respectively. The normalized amplitude of the driving potentials and the driving frequency are fixed at 
$V_p= 0.5$ and $\Omega = 0.01$, while the phase lag between the two time-dependent potentials is set to $\pi/2$. For negative energies, 
the spectral current exhibits nearly equally spaced peaks. As the energy approaches zero and extends to positive values, the peaks 
gradually bunch together, indicating a progressive modification of the energy-resolved transport channels. This behavior is observable 
in both LRH and SRH configurations, although it is more pronounced in the LRH case. Such spectral features clearly signal a crossover 
between adiabatic and non-adiabatic pumping regimes. A quantum pump operates in the adiabatic regime when two conditions are fulfilled. 
First, the electron dwell time, defined as the typical time required for an electron to pass the scattering region, must be much shorter 
than the driving period. In our system, the use of wide-band electrodes (e.g., semi-infinite tight-binding chains with large hopping 
integrals) ensures a short dwell time, so this condition is well satisfied. Second, the energy level spacing, $\Delta$, should significantly
exceed the driving frequency in order to suppress the Floquet-induced level mixing. Although the driving frequency is relatively low ($\Omega=0.01$, corresponding to the GHz range), it is not sufficiently small compared to the energy level spacing across the entire 
energy spectrum. 
\begin{figure}[ht]  
{\centering \resizebox*{8.5cm}{5.5cm}{\hskip -0.5cm \includegraphics{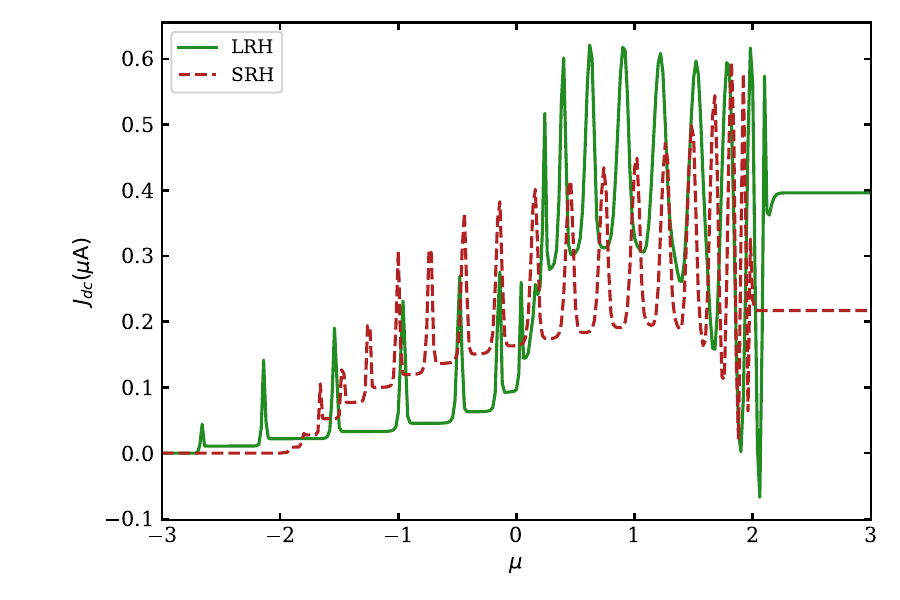}}\par}	
\caption{(Color online). The pumped dc current as a function of normalized chemical potential ($\mu$) for LRH and SRH configurations, 
with $V_p=0.5$, $\Omega =0.01$, and $\delta=\pi/2$.}
\label{fig:3}  
\end{figure}
Consequently, adiabatic and non-adiabatic behaviors can be observed depending on the electron energy. As shown in Fig.~\ref{fig:2}(a), 
the LRH configuration remains predominantly in an adiabatic-like regime for $\omega \lesssim 0$, where the mean level spacing, estimated 
from the spectrum, is $\Delta \approx 0.6$ more than an order of magnitude larger than the driving frequency $(\Omega=0.01)$. At higher 
energies $(\omega\gtrsim 0)$ clear signatures of non-adiabatic transport emerge, manifested by peak bunching and enhanced spectral 
oscillations. In contrast, as depicted in Fig.~\ref{fig:2}(b) the SRH configuration exhibits gradually broadened spectral peaks starting 
from $\omega \gtrsim 0$. Although the peaks remain well separated up to $\omega\approx 2$, indicating relatively weak Floquet-induced 
level mixing, their finite width reduces the degree of adiabatic quantization.

Figure~\ref{fig:3} presents the pumped DC current as a function of the normalized chemical potential $\mu$ for the LRH and SRH 
configurations, with $V_p = 0.5$, $\Omega = 0.01$, and $\delta=\pi/2$.
For the LRH configuration, the pumped DC current exhibits regions where its dependence on the chemical potential becomes relatively 
weak, giving rise to plateau-like features over finite intervals of $\mu$. This behavior, commonly associated with adiabatic charge 
pumping~\cite{Brouwer1998,Thouless1983}, is consistent with the sharp current spectral function structures observed in Fig.~\ref{fig:2}. Importantly, these plateaus are pronounced for $\mu < 0$. However, for $\mu > 0$, where the energy levels become more densely spaced, 
the current loses this robust, flat character. Consequently, outside these plateau regions (particularly for $\mu > 0$ in the LRH case), 
the pumped DC current becomes increasingly sensitive to the chemical potential, signaling a gradual departure from adiabatic transport.
\begin{figure}[ht]  
{\centering \resizebox*{8.5cm}{6cm}{\includegraphics{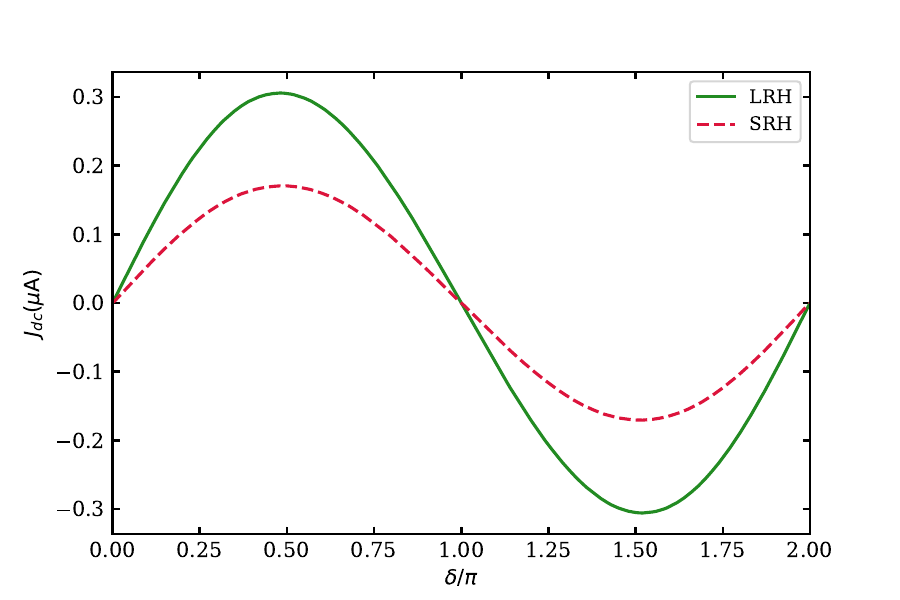}}\par}
\caption{(Color online). The pumped dc current as a function of the phase lag ($\delta$) for LRH and SRH configurations with 
$\mu=0.5$, $V_p=0.5$, and $\Omega = 0.01$.}
\label{fig:4}  
\end{figure}
For the SRH configuration, while the pumped DC current shows structured variations as a function of $\mu$, the closer underlying 
spectral features tend to prevent the formation of clearly defined plateaus across the entire range of $\mu$. Consequently, the current 
remains more sensitive to changes in the chemical potential, indicating reduced robustness of the pumped response compared to the LRH case. 
\begin{figure}[ht]
{\centering \resizebox*{8.5cm}{6cm}{\includegraphics{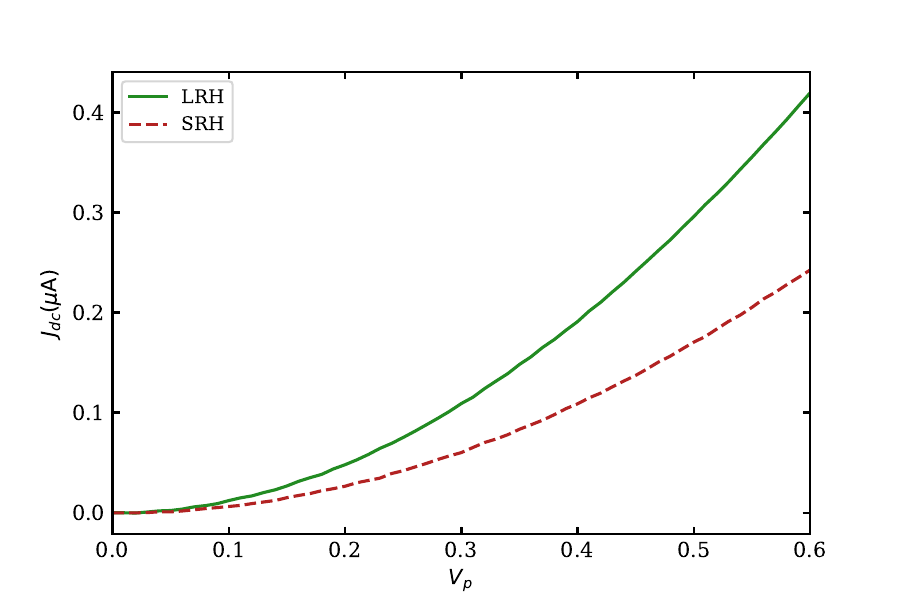}}\par}
\caption{(Color online). The pumped dc current $J_{dc}$ versus the amplitudes of the driving potentials ($V_p$) with $\mu=0.5$ for LRH 
and SRH configurations with $\Omega = 0.01$, and $\delta=\pi/2$.}
\label{fig:5}  
\end{figure}
These distinct transport regimes can be traced back to the fundamental differences in their electronic spectra, imposed by the hopping 
range. In SRH systems, the nearest-neighbor-like coupling yields a relatively symmetric and uniform density of states, leading to the 
absence of robust plateaus in the pumped current.
In contrast, the broken particle-hole symmetry in LRH systems results in a highly asymmetric density of states, where energy levels 
are widely spaced on one side of the spectrum and densely packed on the other~\cite{Sarkar2019}. This directly explains our observations: 
the plateaus in the dc current for $\mu<0$ correspond to the region of well-separated levels enabling adiabatic pumping, while the 
increased sensitivity for $\mu > 0$ stems from the region of closely spaced levels where non-adiabatic effects dominate. Consequently, 
the ability to sustain higher pumped currents in the LRH configuration by tuning $\mu$ originates from this controllable, platform-like 
region in its asymmetric spectrum, a feature absent in the SRH case.

Figure~\ref{fig:4} displays the pumped DC current as a function of the phase lag $\delta/\pi$ between the two time-dependent potentials 
\begin{figure}[ht]
{\centering\resizebox*{8cm}{5cm}{\includegraphics{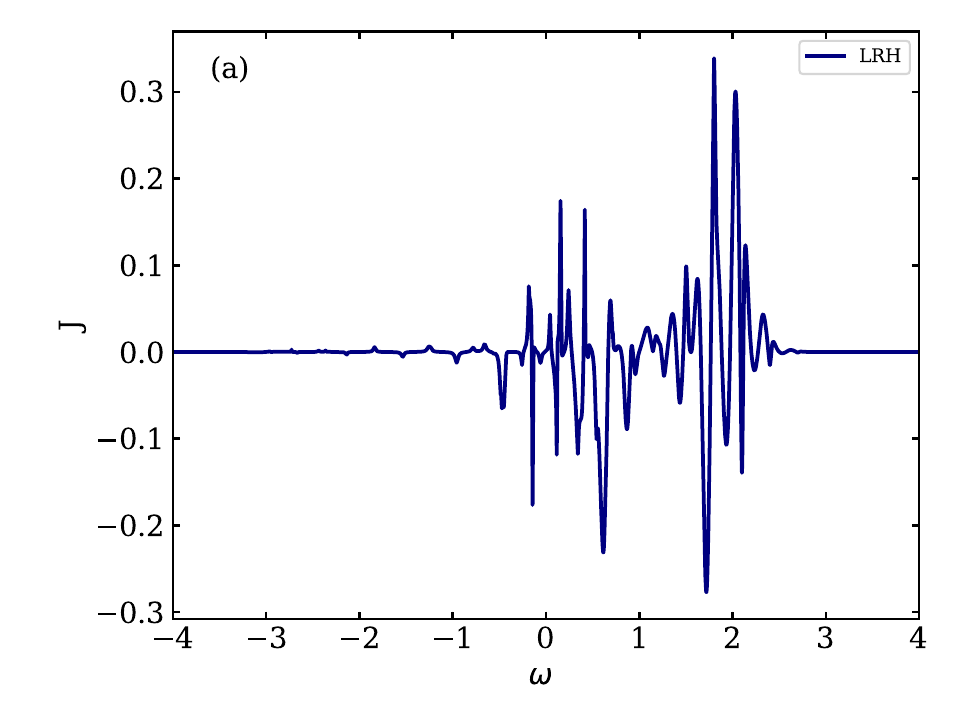}}
\resizebox*{8cm}{5cm}{\includegraphics{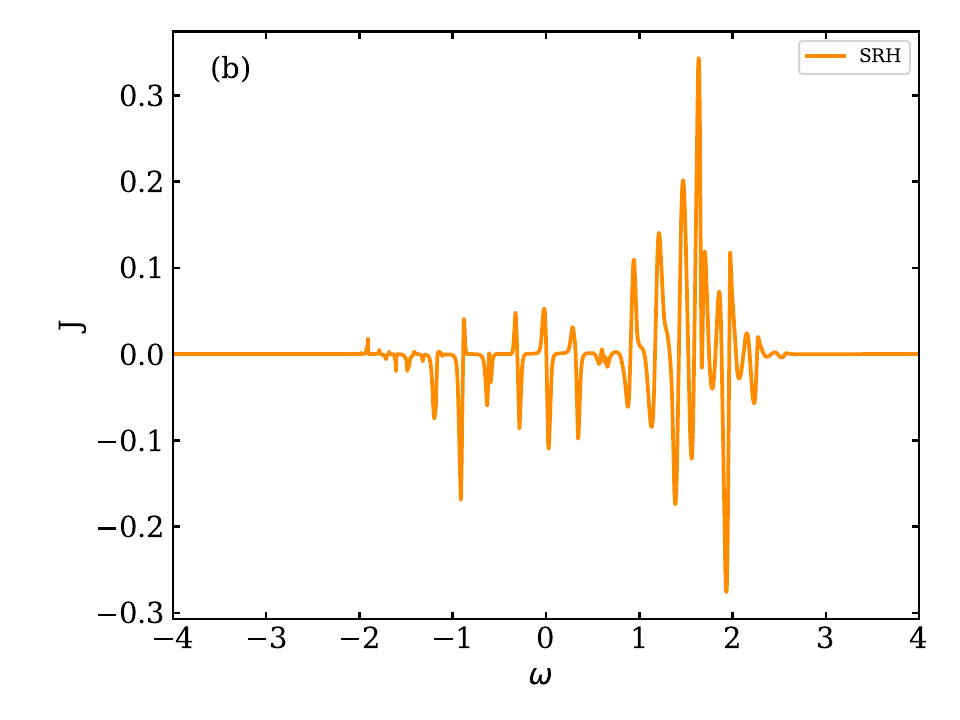}}\par}
\caption{(Color online). The current spectral function versus the energy ($\omega$) for (a) LRH and (b) SRH configurations with 
$V_p = 0.5$, $\Omega = 0.3$, and $\delta=\pi/2$.}
\label{fig:6}
\end{figure}
for the LRH and SRH configurations, evaluated at the chemical potential $\mu=0.5$. All other parameters are identical to those used 
in Fig.~\ref{fig:3}. The pumped current exhibits a clear periodic dependence on $\delta/\pi$, reflecting the periodic nature of the 
driving potentials. In particular, the current vanishes at $\delta=0,\pi,2\pi,\ldots$ for both configurations, indicating that a finite 
phase difference between the two drives is essential to generate a nonzero pumped current.

For weak driving amplitudes in simple chain geometries, the pumped dc current is expected to follow a sinusoidal dependence $J_{\mathrm{dc}}\propto\sin(\delta)$~\cite{Arrachea2005}. In the present case, with a moderate driving amplitude $V_p=0.5$, slight 
deviations from a purely sinusoidal behavior are observed. Such deviations can be attributed to higher-order pumping processes and 
indicate that the system is not strictly confined to the ideal adiabatic limit. 
The behavior of $J_{dc}$ versus the amplitude of the driving potentials ($V_p$) is plotted with $\mu=0.5$ in Fig.~\ref{fig:5} for 
the LRH and SRH configurations. Here, we assume that $\Omega = 0.01$, and $\delta = \pi/2$. For the selected chemical potential, when 
the driving frequency and the amplitudes of the driving potentials are low enough that the system is in the adiabatic regime, the 
pumped current shows a quadratic behavior with respect to the amplitude of the driving potential ($J_{dc}\propto V_p^2$), which is 
in agreement with the literature~\cite{Arrachea2005}. 

Figures~\ref{fig:6}(a) and \ref{fig:6}(b) show the current spectral function $(j)$ versus the electron energy $\omega$ for the LRH and 
SRH cases, respectively. The normalized amplitude of the driving potentials is set to $V_p = 0.5$, with a phase lag $\delta = \pi/2$. 
The driving frequency is chosen as $\Omega = 0.3$, which is sufficiently large to drive the system well beyond the adiabatic regime 
over the entire energy range. At such high frequencies, the time-dependent modulation induces strong Floquet side-band mixing, leading 
to the opening of multiple resonant transport channels. As a result, the current spectral function becomes significantly broadened in 
energy for both configurations, in contrast to the low-frequency case discussed earlier. While both LRH and SRH systems exhibit pronounced
spectral oscillations, the detailed spectral patterns differ between the two configurations, reflecting the distinct hopping ranges involved. 

\begin{figure}[ht]  
{\centering \resizebox*{8.5cm}{5.5cm}{\includegraphics{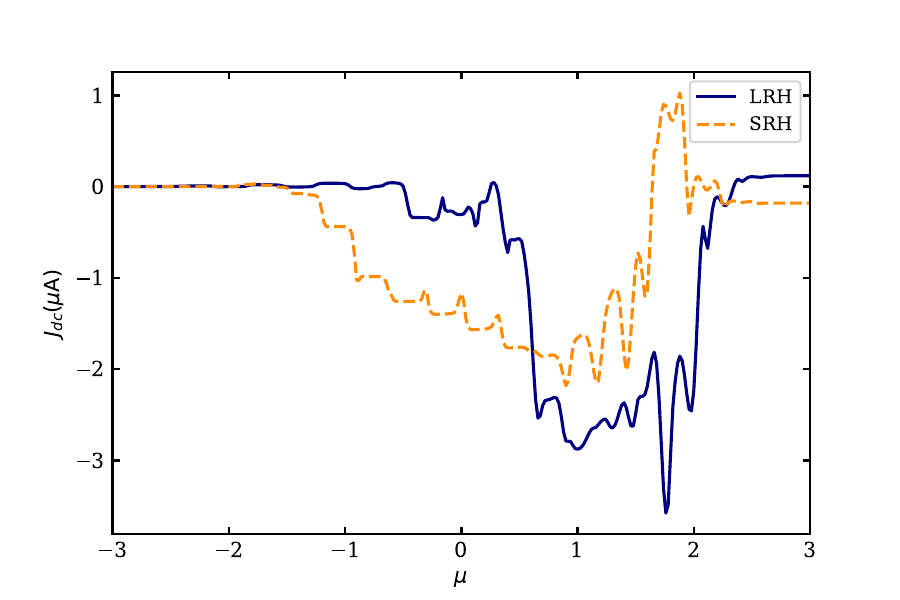}}\par}
\caption{(Color online). The pumped dc current ($J_{dc}$) versus $\mu$,  for the LRH and SRH regimes for $V_p$=0.5, $\Omega = 0.3$, 
and $\delta$=$\pi/2$.}
\label{fig:7}  
\end{figure}

Figure~\ref{fig:7} shows the pumped DC current as a function of the normalized chemical potential ($\mu$) for the LRH and SRH helices 
at $\Omega = 0.3$, with $V_p = 0.5$ and $\delta = \pi/2$. For the LRH configuration, the pumped DC current exhibits pronounced oscillatory
behavior as the chemical potential is varied, indicating a strong sensitivity to $\mu$ over the entire range considered. In contrast to 
the plateau-like features observed at lower frequencies, the current at this driving frequency does not display extended regions of weak 
$\mu$ dependence, but instead shows sharp extrema and rapid variations, particularly for $\mu>0$. For the SRH configuration, while the 
pumped DC current also exhibits structured variations as a function of $\mu$, the reduced separation of the underlying spectral features 
leads to a smoother response with a smaller current amplitude compared to the LRH case. Consequently, at $\Omega = 0.3$, the pumped 
current in the SRH configuration remains less pronounced and more gradually varying than in the LRH configuration.

A comparison between the low- and high-frequency regimes reveals the influence of the driving frequency on the pumping characteristics. 
At $\Omega = 0.01$, the pumped current, particularly for the LRH configuration, displays relatively flat regions as a function of $\mu$ 
(see Fig.~\ref{fig:3}), consistent with an approximately adiabatic response. When the frequency is increased to $\Omega = 0.3$, these 
flat regions are no longer observed and the current becomes more oscillatory and sensitive to the chemical potential. This behavior 
suggests a gradual crossover from an adiabatic to a more dynamical pumping regime as the driving frequency increases.      
\begin{figure}[ht]  
{\centering \resizebox*{8.5cm}{5.5cm}{\includegraphics{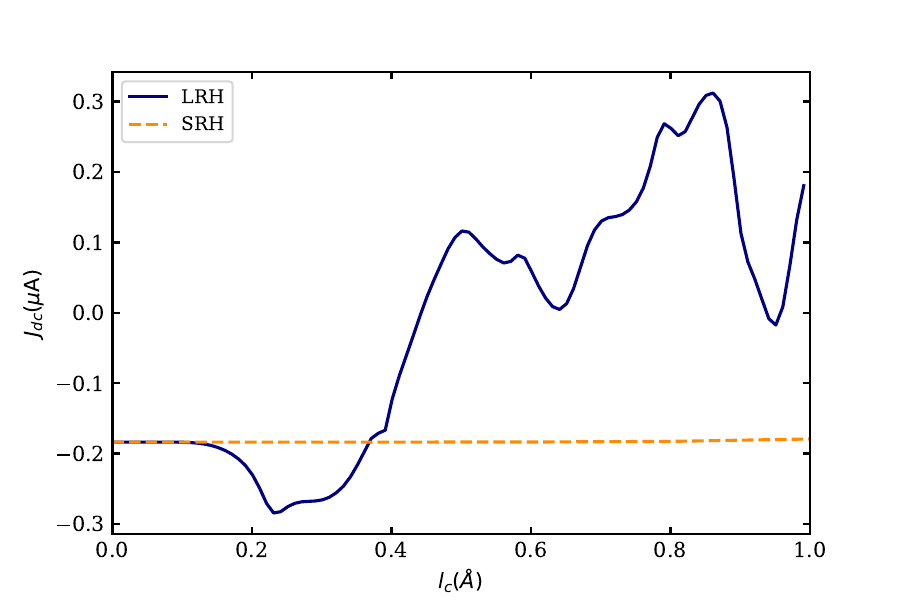}}\par}
\caption{(Color online). The pumped dc current as a function of the decay constant ($\ell_c$) for LRH and SRH configurations with 
$\mu=3$, $V_p$=0.5, $\Omega = 0.3$, and $\delta$=$\pi/2$ }
\label{fig:8}
\end{figure}
To explore the effect of the helix structure on the pumped dc current, we focus on a key geometrical parameter: the decay constant 
($\ell_c$). As $\ell_c$ increases, the contribution of next-nearest neighbors to the hopping term gradually grows, and the system 
evolves from SRH-like to LRH behavior. Figure~\ref{fig:8} shows the pumped dc current as a function of the decay constant ($\ell_c$) 
for both configurations. The normalized chemical potential of the reservoirs is set to $\mu=3$, and other pumping parameters are the 
same as in Fig.~\ref{fig:7}. For small $\ell_c$, the LRH reduces to SRH-like behavior and both configurations exhibit the same pumped 
DC currents. As $\ell_c$ increases long-range hopping becomes significant in the LRH configuration, resulting in non-monotonic current 
varying between $-0.3$ and $+0.3$ $\mu A$. In contrast, the SRH current remains almost unaffected, as only nearest-neighbor hopping 
contributes. These results indicate that in the LRH configuration the helix structure can serve as an effective control parameter 
for the pumped DC current.  

\section{Summary and outlook}

We have studied quantum charge pumping in single-stranded helical systems under periodic driving, comparing long-range hopping (LRH) 
and short-range hopping (SRH) using the Keldysh non-equilibrium Green's function formalism. Our results (see Figs.~\ref{fig:3}-\ref{fig:7}) 
show that the hopping range strongly influences both the spectral and pumped DC currents.

In the LRH configuration, the DC current exhibits plateau-like regions for chemical potentials corresponding to sparsely spaced energy 
levels, consistent with adiabatic pumping. These plateaus diminish as the energy levels become more densely spaced or when the driving 
frequency increases, reflecting the gradual onset of non-adiabatic effects. 

The phase dependence of the pumped current is approximately sinusoidal for moderate driving amplitudes, vanishing at 
$\delta = 0, \pi, 2\pi, \ldots$ for both LRH and SRH cases, consistent with the literature~\cite{Arrachea2005}. 

The most striking result of this work, shown in Fig.~\ref{fig:8}, concerns the role of the helix structure. Remarkably, the hopping 
decay exponent $\ell_c$ serves as a control parameter that can modify both the magnitude and sign of the pumped current in LRH systems, 
spanning the range $-0.3$ to $+0.3$ $\mu A$. In sharp contrast, the SRH current remains entirely unaffected by $\ell_c$, as only
nearest-neighbor hopping contributes. This key distinction demonstrates that the helix structure enables tunable charge pumping 
in the presence of long-range hopping.

These findings indicate that the interplay between hopping range, driving conditions, and energy spectra governs the characteristics 
of charge pumping in helical systems.

\bibliographystyle{apsrev4-2}
\bibliography{references}

\end{document}